\documentclass[prd,aps,floats,preprint,superscriptaddress,nofootinbib,tightenlines]{revtex4}
\usepackage{epsfig}
\usepackage{amsmath}

\def\Dsl{\hbox{/\kern-.6000em D}} %roman D

\def\dsl{\,\raise.15ex\hbox{/}\mkern-13.5mu D}

\def\ltap{\ \raise.3ex\hbox{$<$\kern-.75em\lower1ex\hbox{$\sim$}}\ }
\def\gtap{\ \raise.3ex\hbox{$>$\kern-.75em\lower1ex\hbox{$\sim$}}\ }
\def\OMIT#1{}

\def\OMIT#1{}

\newcommand{\nn}{\nonumber}

\newcommand{\bea}{\begin{eqnarray}}
\newcommand{\eea}{\end{eqnarray}}

\begin{document}
\title{On Non-Relativistic Conformal Field Theory and Trapped Atoms:
Virial Theorems and the  State-Operator Correspondence in Three Dimensions}

\author{Thomas Mehen}
\email{mehen@phy.duke.edu}
\affiliation{Department of Physics, Duke University, Durham NC 27708, USA}
\date{\today}
\vspace{0.5in}
\begin{abstract}

The field theory of nonrelativistic fermions interacting via contact interactions  
can be used to calculate the properties of few-body systems of cold atoms 
confined in harmonic traps.  The state-operator correspondence of Non-Relativistic 
Conformal Field Theory (NRCFT) shows that the energy eigenvalues (in oscillator units) 
of $N$ harmonically trapped fermions can be calculated from the scaling dimensions of $N$-fermion 
operators in the NRCFT. They are also in one-to-one correspondence with zero-energy,
scale-invariant solutions to the $N$-body problem in free space. We show that these
two mappings of the trapped fermion problem to  free space problems are related by an automorphism
of the $SL(2,R)$ algebra of the conformal symmetry of fermions at the unitary limit. 
This automorphism exchanges the internal Hamiltonian of the gas with the trapping potential and hence 
provides a novel method for deriving  virial theorems for trapped Fermi gases at the unitary limit.
We also show that the state-operator correspondence can be applied directly 
in three spatial dimensions by calculating the scaling dimensions of two- and three-fermion 
operators and finding agreement with known exact results for energy levels 
of  two and three trapped fermions at the unitary limit.

\end{abstract}

\maketitle

\section{Introduction}

The problem of few-body atomic interactions in the presence of 
external confining potentials is motivated by recent advances in experimental
atomic physics as well as theory. Recent experiments have realized optical 
lattices with two confined in a potential well~\cite{smgke,vsbhdr,twlsg,oohesb}.
Such atomic states have been proposed for implementing quantum logic gates~\cite{bcjd,jea,bdw,dd}.
Theoretically, the problem of two atoms interacting
via short-range forces has been solved  in Ref.~\cite{Busch:1999}, see also Refs.\cite{Tiesinga:2000,
Blume:2002,Block:2002,Bolda:2002,Stock:2005,Idziaszek:2006a,Idziaszek:2006b}. 
An experimental confirmation of the prediction for the
ground state energy of two trapped atoms 
as a function of scattering length was recently performed in Ref.~\cite{smgke}.
If the scattering length 
of the atoms is tuned to infinity and effective range terms are neglected then 
the S-wave scattering cross section is $4\pi/p^2$ where $p$ is the relative momentum.
This cross section is at the upper limit allowed by unitarity. The quantum mechanics 
problem of three particles at the unitary limit in the presence of a harmonic 
potential has been solved in Ref.~\cite{wc3}, see also Ref.~\cite{jhp}. 
 
An outstanding open problem is the many-body problem of fermions at unitary,
which has been investigated by numerous authors using a wide variety of methods.
Gases of trapped fermions whose interactions have been tuned to the unitary 
limit by means of a Feshbach resonance have been realized experimentally~\cite{jet,reg,bou}.
For a review of experimental and theoretical results, see Refs.~\cite{yc,gps}.
Both the homogeneous unitary Fermi gas as well as the unitary Fermi gas in the 
presence of harmonic traps are clearly of interest. Though the many-body 
physics problem presents physical challenges not present in the two- and three-body
problems, the existence of exact solutions for $N=2$ and $3$, where $N$ is the number of fermions,
can provide important inputs for the case of arbitrary $N$. For example,  Ref.~\cite{Papenbrock:2005bd}
proposes a scale-invariant density functional for the unitary Fermi gas 
whose parameters are fixed by matching the known analytic solutions for two fermions
at the unitary limit in the harmonic trap. Corrections due to a finite scattering length 
and effective range are included in Ref.~\cite{Bhattacharyya:2006fg}.
This provides another motivation for studying few-body 
trapped fermion problems.

An interesting theoretical development is the state-operator correspondence  which relates the problem of finding
the energy eigenvalues of $N$ trapped fermions at the unitary limit  to the problem of  finding the scaling
dimensions of primary operators in  a Non-Relativistic Conformal Field Theory (NRCFT)~\cite{Nishida:2007pj}. Note
that the NRCFT is defined in the absence  of an external potential, so the state-operator correspondence  relates 
a property of the theory of $N$ fermions in  free space to the properties of $N$ trapped fermions. Another mapping of the trapped
$N$-fermion problem to the free space $N$-fermion problem is derived in Ref.~\cite{wc1}. These authors map the
problem of  harmonically trapped fermions at the unitary limit to the problem of finding zero-energy, scale
invariant  eigenfunctions of the $N$-body problem in the absence of any external potential.  One goal of this paper
is to better understand the relationship between these two mappings.

The other main goal of this paper is to show how the
state-operator correspondence can be applied directly in three dimensions.
For two spatial dimensions ($d=2$), the theory of fermions at the unitary limit is equivalent 
to noninteracting fermions while in $d=4$ the theory is equivalent to noninteracting bosons~\cite{nussinov2}.
Therefore, in $2+\epsilon$ dimensions and $4-\epsilon$ dimensions a perturbation theory 
in $\epsilon$ can be used to analyze the properties of unitary fermions~\cite{Nishida:2006br,Nishida:2006eu}.
In Ref.~\cite{Nishida:2007pj}, the $\epsilon$ expansion is combined with the state-operator
correspondence to calculate the energy levels of few-body atomic systems in harmonic traps.
Operator scaling dimensions are calculated in a perturbative series in $\epsilon$ and 
Pad\'e approximants are used to interpolate between $d=2$ and $d=4$ to obtain results for the
most physically interesting case of $d=3$. In this paper, we will illustrate how the state-operator
correspondence can be applied directly in $d=3$. Using a low energy effective field theory
for two component fermions interacting via $S$-wave contact interactions,
we calculate the scaling dimensions of $S$-wave $N=2$ and $N=3$ fermion  operators and 
find agreement with the exact solutions for the energy levels of two trapped fermions,
as well as the lowest energy state of three trapped fermions in an $S$-wave. 

The low energy interactions of few-body systems can be studied using the methods of effective field
theory. These methods are useful when the typical momentum times  the range of the interactions is
much less than one. This is 
the case for cold atoms, where a complete model for the potential is not required and many quantities can be 
computed in terms of the $S$-wave scattering length alone. At these energies the $S$-wave scattering amplitude
for two fermions with momentum $\pm p$ is 
\bea
{\cal A} &=&\frac{4\pi}{M} \frac{1}{p \cot \delta(p) - i p} \nn \\
&=& \frac{4\pi}{M} \frac{1}{-1/a + r_0 \,p^2/2 + ... - i p} \, , \nn
\eea
where $a$ is the scattering length and $r_0$ is the effective range. In the limit $r_0 \,p \ll 1$,  
effective range corrections can be neglected and the two particle scattering amplitude 
is exactly reproduced by a nonrelativistic field theory with a single $S$-wave contact interaction.
The Lagrangian for this nonrelativistic field theory is 
\bea\label{lag}
{\cal L} = \psi^\dagger \left(i \partial_t + \frac{\nabla^2}{2 M}\right)\psi -
\frac{C_0(\mu)}{4}\psi^\dagger \psi^\dagger \psi \psi\, ,
\eea
where $\psi$ is a two-component field operator that annihilates fermion quanta. Here 
$\psi \psi = \epsilon^{\alpha \beta} \psi_\alpha \psi_\beta$, so scattering occurs 
in the $S$-wave, spin-singlet channel only. The coupling constant, $C_0(\mu)$, is given by
\bea\label{cc}
C_0(\mu) = \frac{4\pi}{M}\frac{1}{-\mu+1/a} \, ,
\eea
where $\mu$ is the dimensional regularization (DR) parameter and  Power Divergence 
Subtraction (PDS) scheme is used to regulate loop integrals~\cite{Kaplan:1998tg}. In DR, loop integrals which are linearly divergent when regulated 
with a cutoff can become finite because DR discards power law divergences. These can be restored within the framework 
of DR by subtracting poles in one lower dimension, then the DR parameter, $\mu$, enters the calculation
of loop integrals in the same way that a hard cutoff would. 
In the Minimal Subtraction (MS) scheme, where the
linear divergences are discarded, $C_0 = 4\pi a/M$. Thus, it is clear that to make
sense of the Lagrangian in the limit $a\to\pm\infty$, one needs to use a hard cutoff, PDS, or some
other regularization scheme that keeps track of linear  divergences. We will see below that in order
to obtain scaling dimension of operators that are consistent with the state-operator correspondence of
NRCFT, we must also use one of these schemes.

Since the work of Ref.~\cite{Mehen:1999nd}, it is known that the $a\to \pm \infty$ limit of  
the  theory of two-component fermions in Eq.~(\ref{lag}) is conformally invariant.~\footnote{For 
bosons or fermions with more than two degrees of freedom, an $S$-wave three-body contact interaction
is relevant and violates scale invariance~\cite{Bedaque:1998kg}.} The nonrelativistic 
scale transformation  is
\bea\label{scale}
\vec{x}^{\,\prime} = \lambda \,\vec{x} \qquad t^\prime = \lambda^2\, t 
\qquad \psi^\prime(\vec{x}^{\,\prime},t^\prime) = \lambda^{-3/2} \psi(\vec{x}, t) \, ,
\eea
and the nonrelativistic conformal transformation is 
\bea\label{conformal}
\vec{x}^{\, \prime} = \frac{\vec{x}}{1+c \,t } \qquad t^\prime = \frac{t}{1+c\, t} \qquad
\psi^\prime(\vec{x}^{\, \prime},t^\prime) = (1+c \,t)^{3/2} 
 \exp\left(\frac{-i M \, c \, \vec{x}^{\,2}}{2(1+c \,t)}\right)
\psi\left(\vec{x},t \right)\,. 
\eea
It is straightforward to show that the scale and conformal transformations are symmetries of the non-interacting theory.
For generic values of $a$ the contact interaction in Eq.~(\ref{lag}) breaks these symmetries.
Since the two-particle $S$-wave cross section is independent of any scale when $a\to \pm \infty$,
it is natural to expect the theory to be scale and conformally invariant in this limit.
Ref.~\cite{Mehen:1999nd} showed that the off-shell $2\to 2$ scattering amplitude calculated in the
theory of Eq.~(\ref{lag}) is invariant under the Ward identities implied by scale and conformal
transformations when  $a\to \pm\infty$. Ref.~\cite{Nishida:2007pj} gives a simple
argument for why any particle number conserving theory that is scale invariant
should also be  invariant under conformal transformations. 

When $a\to \pm \infty$, the Hamiltonian, $H$, the generator of nonrelativistic scale transformations, $D$, and the generator
of conformal transformations, $C$, form  an $SL(2,R)$ algebra. We will show below that the two mappings of the trapped fermion
problem to  free space fermion problems that were discussed earlier are related by an automorphism of the $SL(2,R)$ group.
This automorphism exchanges the generators $C$ and $H$. Since the generator $C$ is just the external potential for the trapped
fermions, this automorphism interchanges the trapping potential and the internal Hamiltonian of the gas. Therefore, the
automorphism can be used to provide a novel group theoretical derivation of virial theorems for trapped Fermi gases
at the unitary limit. A virial theorem was first derived using the assumption of universality, the local density approximation, and thermodynamic
arguments in Ref.~\cite{tkt}. The virial theorem was then rederived and generalized using the wavefunctions of the
pseudopotential model  of the unitary Fermi gas in Ref.~\cite{wc1}. Another derivation of the virial theorem using the
Hellmann-Feynman theorem appears in Ref.~\cite{son3}. Our derivation is novel in that the approach is group theoretical and
relies only on the $SL(2,R)$ algebra. The virial theorems hold for $N$-body energy eigenstates as well for thermal ensembles,
and can be applied to spin polarized or unpolarized gases.

The paper is organized as follows: in the next section, we review basic facts about NRCFT's
and the state-operator correspondence, as well as the correspondence of Ref.~\cite{wc1} which 
relates eigenstates of trapped fermions to zero-energy, scale-invariant eigenfunctions in free space.
We discuss the automorphism of the $SL(2,R)$ algebra which relates these mappings and show how it can be used
to derive the virial theorems. In section III, we derive the scaling dimension of  operators with $N=2$
and $N=3$ and show that these agree with analytic results for the  energies 
of trapped fermions. In Section IV, we conclude. In the Appendix, we solve the problem
of two trapped atoms with arbitrary short-range interactions. This 
was  first done for arbitrary scattering length in Ref.~\cite{Busch:1999} using the method
of pseudopotentials. Here we solve the problem by calculating Green's functions for two particles in the 
trap using the field theory of Eq.~(\ref{lag}).

\section{NRCFT, $SL(2,R)$ Automorphisms, and Virial Theorems}

In this section we begin by reviewing NRCFT and the two mappings of the problem of trapped fermions at the unitary
limit to free space problems~\cite{Nishida:2007pj,wc1}. We then show that these two mappings are related by an
automorphism of the $SL(2,R)$ conformal symmetry algebra. This is the main result of this section. 
The automorphism is realized by a unitary transformation, generated by the Hamiltonian of the trapped fermions, that
interchanges the internal Hamiltonian of the Fermi gas with the external trapping potential. This unitary
transformation can then
be used to provide a simple derivation of the virial theorems for trapped fermions at the unitary limit.

The many-body Hamiltonian for harmonically trapped fermions in second quantized form is
the sum of an internal Hamiltonian, $H_{\rm int}$, and an external potential, $V_{\rm ext}$,
which are given by
\bea
H_{\rm int}&=& \int d^3x \, \left[ \psi^\dagger \left(-\frac{\nabla^2}{2 M}\right)\psi +
\frac{C_0(\mu)}{4}\,\psi^\dagger \psi^\dagger \psi \psi\, \right] \nn \\
V_{\rm ext} &=& \int d^3x\, \frac{1}{2} M \, \omega^2 \, \vec{x}^{\,2} \, \psi^\dagger \psi \, .
\eea
After the following rescaling,
\bea
\vec{x} \to \frac{\vec{x}}{\sqrt{M\omega}} \qquad
\psi \to (M \omega)^{3/4} \psi  \qquad 
\mu \to \sqrt{M \omega} \, \mu \qquad 
a \to \frac{a}{\sqrt{M\omega}} \, ,
\eea
which renders all these quantities dimensionless (we are using $\hbar =1$ units),
we find
\bea
H_{\rm int} &=& \omega  \int d^3x \, \left(\psi^\dagger \left(-\frac{\nabla^2}{2}\right)\psi +
\frac{\hat{C}_0(\mu)}{4}\,\psi^\dagger \psi^\dagger \psi \psi\, \right) \nn \\
&\equiv& \omega \, H \nn \\
V_{\rm ext} &=& \omega \int d^3x \,\frac{1}{2} \,\vec{x}^{\,2} \, \psi^\dagger \psi \nn \\
&\equiv& \omega \, C \, .
\eea 
Here we have defined $\hat{C_0}(\mu) = M \,C_0(\mu)$, so that $\hat C_0(\mu)$ is independent of $M$. 
This shows that we can set $M = \omega =1$ and measure all energies in units of
the fundamental oscillator energy, $\omega$. Lengths are measured in units of $a_{\rm osc} =1/\sqrt{M \omega}$.
In this section, we will use these units and work with  $H$ and $C$ rather than $H_{\rm int}$
and $V_{\rm ext}$. $C$ is the generator of conformal transformations~\cite{Nishida:2007pj}.

If we modify the definition of the scale transformation to include the appropriate transformation 
on $\mu$, 
\bea\label{scalemod}
\vec{x}^{\,\prime} = \lambda \,\vec{x} \qquad t^\prime = \lambda^2\, t \qquad \mu^\prime = \lambda^{-1} \mu \qquad 
\psi^\prime(\vec{x}^{\,\prime}, t^\prime) = \lambda^{-3/2}\, \psi(\vec{x},t)\, ,
\eea
we find 
\bea
H^\prime = \lambda^{-2} H \, ,
\eea
when $a=\pm\infty$. Though the scale transformation of Eq.~(\ref{scalemod}) differs from that of Eq.~(\ref{scale}) by additional
rescaling of $\mu$, the Ward identities derived in Ref.~\cite{Mehen:1999nd} will still hold for any renormalized 
Green's function that is $\mu$ independent. Likewise, to see the conformal invariance of 
the theory defined by Eq.~(\ref{lag}) explicitly, one must modify the conformal transformation
in Eq.~(\ref{conformal}) to include a time-dependent rescaling of $\mu$.
If $D$ is the generator of scale transformations then 
\bea
H^\prime = e^{i \alpha D} H e^{-i \alpha D} = e^{-2 \alpha} H \qquad (\alpha = \log \lambda) \, ,
\eea
which gives the commutation relation $[D,H] = 2 i H$.

In a NRCFT, the Hamiltonian, $H$, dilatation operator, $D$,
and conformal generator, $C$, obey the following commutation relations:
\bea\label{sl2R}
&a)& \qquad [D, H] = 2 i H \nn \\
&b)& \qquad [D, C] = - 2 i C\nn \\
&c)& \qquad [H, C] = - i D \, ,
\eea
which are the commutation relations of the group $SL(2,R)$. For the theory of Eq.~(\ref{lag}), we have given 
$H$ and $C$ above and $D$ is given by\footnote{The explicit expressions 
for $C$ and $D$ are valid at $t=0$. For arbitrary $t$, 
$C(t)=C(0) + t^2 H - t D(0)$ and $D(t) = D(0)-2 t H$. The explicit
time dependence can be fixed by requiring $\dot{A}(t) = -i[A,H] +\partial A/\partial =0, A =C,D$, 
 which is required for conserved charges 
that generate a symmetry of the Hamiltonian. See Ref.~\cite{de Alfaro:1976je} for a one-dimensional
conformally invariant quantum mechanical system.}
\bea\label{DandC}
D &=& \int d^3x \, \vec{x} \cdot \psi^\dagger \left(-\frac{i}{2}\overleftrightarrow{\nabla} \right) \psi 
= \int d^3x \, \vec{x} \cdot \vec{j}(x) \, .
\eea
where $\vec{j}(x)$ is the particle current density. Eq.~(\ref{sl2R}b) follows
automatically from the definitions of $D$ and $C$ and the equal time commutation relations of 
$\psi$ and $\psi^\dagger$. Eq.~(\ref{sl2R}c) is actually quite general
and will hold for any theory in which particle number is locally conserved. Note that
$C = \int d^3x \,\frac{1}{2} \vec{x}^{\,2} \,n(x)$, where $n(x)$ is the particle density operator. 
The commutator of the Hamiltonian is proportional to the divergence of the particle current~\cite{Nishida:2007pj} 
\bea\label{currentcons}
[H, n(x)] = -i \partial_t n(x) = i \vec{\nabla} \cdot \vec{j}(x) \, ,
\eea
due to current conservation. Eq.~(\ref{sl2R}b) follows by multiplying Eq.~(\ref{currentcons}) by
$\vec{x}^{\,2}/2$ and integrating over all space. So if Eq.~(\ref{sl2R}a), which is 
the requirement of scale invariance, is satisfied in a particle number conserving theory
then the theory will also be invariant under the full $SL(2,R)$ conformal group.
To complete the algebra of the Schr\"odinger group (the largest space-time 
symmetry group of free nonrelativistic quantum mechanics), we also need the commutation relations of 
$H$, $D$, and $C$ with 
other symmetry generators: momentum, $\vec{P}$, angular momentum, $\vec{J}$, 
Galilean boosts, $\vec{K}$, and particle number, $N$. 
The nonvanishing commutators involving $K_i$, $P_i$, $D$, and $N$ are:
\bea
\, [K_i,P_j] = i \,N \delta_{ij} \qquad 
[D,P_i] = i P_i \qquad 
[D,K_j] = -i K_i \, .
\eea
Commutation relations involving $\vec{J}$ are easily deduced from rotational invariance.

Primary operators in the NRCFT are defined by ${\cal O} \equiv {\cal O}(\vec{x}=0,t=0)$ and
\bea
[\vec{K},{\cal O}] = [C,{\cal O}]= 0. 
\eea
Note that primary operators are defined to be located at the origin of space and time.
The particle number ($N_{\cal O}$) and scaling dimension ($\Delta_{\cal O}$) 
of the primary operator are defined by
\bea
\, [D,{\cal O}] &=& i \Delta_{\cal O} {\cal O} \, ,\nn \\
\, [N,{\cal O}] &=&  N_{\cal O} {\cal O} \, .
\eea

If we translate the primary operator ${\cal O}$ to another point in space-time
\bea
{\cal O}(\vec{x},t) = e^{i H t - i \vec{P}\cdot \vec{x}} {\cal O}(0) e^{-i H t + i \vec{P}\cdot \vec{x}} \, ,
\eea
then it is straightforward to show using the commutation relations listed above 
that
\bea
\, [\vec{K},{\cal O}] &=&  (-i t \partial_i +N_{\cal O} x_i) {\cal O}\nn \\
\, [C,{\cal O}] &=& -i( t^2 \partial_t +t \vec{x}\cdot\vec{\partial} +t \Delta_{\cal O}){\cal O}
+\frac{\vec{x}^{\,2}}{2} \,N_{\cal O} {\cal O} .
\eea
The field $\psi$ has $N_\psi = -1$ and $\Delta_\psi=d/2$, where $d$ is the dimensionality
of space. The density operator, $\psi^\dagger \psi$, has  $N_{\psi^\dagger \psi} = 0$ and 
$\Delta_{\psi^\dagger \psi} = d$. For a finite conformal transformation we have
\bea
{\cal O}^\prime(\vec{x},t) &=& e^{-i \lambda C} {\cal O}(\vec{x},t) 
e^{i \lambda C} \nn \\
&=&
\frac{1}{(1+\lambda \, t)^{\Delta_{\cal O}}}
\exp\left(\frac{-i N_{\cal O} \,\lambda \, \vec{x}^{\,2}}{2(1+\lambda \, t)}\right) 
{\cal O}\left(\frac{\vec{x}}{1+\lambda \, t},\frac{t}{1+\lambda \,t}\right)\, .
\eea
which agrees with Eq.~(\ref{conformal}) for the case ${\cal O}= \psi$. 

Next we discuss 
consequences following from the algebra in Eq.~(\ref{sl2R}). Let us define
\bea
H_{\rm osc} &\equiv& H + C \nn \\
L_{\pm} &=& H-C \pm i D \,.
\eea
The $L_\pm$ are ladder operators that raise and lower energy eigenvalues of $H_{\rm osc}$ by two oscillator
units, as can be seen 
from the commutation relations
\bea
\, [L_\pm,H_{\rm osc}] &=& \mp \,2 L_\pm \nn \\ 
\, [L_-,L_+] &=& 4 H_{\rm osc} \, ,
\eea
which are easily derived from Eq.~(\ref{sl2R}). Eigenstates of $H_{\rm osc}$  come in 
infinite towers of equally spaced states. The ground state of one of these towers is denoted 
by $|\psi_0\rangle$ which satisfies $L_-|\psi_0\rangle =0$. 

The problem of finding energy eigenstates for the trapped particles can be mapped to the
free space theory in one of two ways. The method of Ref.~\cite{Nishida:2007pj} begins by noting 
that
\bea\label{rel1}
e^H L_- e^{-H} = - C \, ,
\eea
and furthermore 
\bea
C {\cal O}^\dagger|0\rangle &=& [C, {\cal O}^\dagger]|0\rangle \nn \\
&=& 0 \, ,
\eea
where ${\cal O}$ is a primary operator and $|0\rangle$ is the vacuum. Then
\bea
L_- e^{-H}{\cal O}^\dagger|0\rangle = -e^{-H} C {\cal O}^\dagger |0\rangle = 0 \, 
\eea
so the ground state of the tower is $|\psi_0\rangle = e^{-H}{\cal O}^\dagger|0\rangle$.
We find that from any primary operator we can construct a tower of eigenstates of $H_{\rm osc}$.
It is straightforward to show that~\cite{Nishida:2007pj}
\bea
H_{\rm osc} e^{-H}{\cal O}^\dagger|0\rangle = e^{-H}(C-i D) {\cal O}^\dagger |0\rangle 
= \Delta_{\cal O} e^{-H}{\cal O}^\dagger|0\rangle \, .
\eea
Thus the scaling dimension of the operator in the NRCFT (in the absence of an external potential)
gives  the ground state energy (in oscillator units) of the corresponding state $|\psi_0\rangle$
in the problem with an external harmonic potential.

Ref.~\cite{wc1} pointed out another correspondence between eigenstates of the trapped fermions,
$H_{\rm osc}$, and zero-energy, scale invariant eigenstates of $H$. The result of Ref.~\cite{wc1}
can be obtained starting with a relation analogous to Eq.~(\ref{rel1}),
\bea\label{rel2}
e^C L_- e^{-C} = H \, .
\eea
From this relation it is clear that $|\psi_0\rangle = e^{-C}|\psi_\nu\rangle$, where
$|\psi_\nu\rangle$ is a zero-energy eigenstate of the Hamiltonian, $H |\psi_\nu\rangle =0$.
For this state to be an eigenstate of $H_{\rm osc}$, it must be an eigenstate
of $i D$ as well,
\bea 
i D |\psi_\nu\rangle = \left(\nu +\frac{3}{2}N\right) |\psi_\nu \rangle \, .
\eea
 The energy eigenvalue of
$|\psi_0\rangle$ is 
\bea\label{enev}
  H_{\rm osc} |\psi_0\rangle &=& e^{-C} \,e^C H_{\rm osc}\, e^{-C}|\psi_\nu\rangle \nn \\
&=& e^{-C} (H+i D)|\psi_\nu\rangle \nn \\
 &=& \left(\nu +\frac{3}{2}N \right)|\psi_0\rangle \,. 
\eea
To understand the significance of  $\nu$, note that the $N$-body wavefunction associated with
the state $|\psi_\nu\rangle$ is 
\bea
\psi_\nu(\vec{x}_i) =\langle 0 | \prod_{i=1}^N \psi(\vec{x_i}) |\psi_\nu\rangle \,  .
\eea
Using 
\bea
e^{-i  \alpha D}\psi(\vec{x_i}) e^{i \alpha D} = e^{\frac{3}{2} \alpha}\psi(e^\alpha \vec{x}_i)
\eea
 it is easily seen that 
\bea
\psi_\nu\left(\frac{\vec{x}_i}{\Lambda}\right)  =  \Lambda^{-\nu}\psi_\nu(\vec{x}_i) \, ,
\eea
so the $N$-body wavefunction for the state $|\psi_\nu\rangle$ is a homogeneous function
of the $N$-body coordinates. Note that the $N$-body wavefunction for the trapped problem is given by
\bea
\langle \vec{x}_i|\psi_0\rangle &=& \langle \vec{x}_i | e^{-C} |\psi_\nu\rangle
\nn \\
&=& e^{-\sum_i \vec{x}_i^2/2} \psi_\nu(\vec{x}_i) \, .
\eea

To understand the relationship between the two mappings of the trapped problem
to free space problems, we observe that
Eqs.~(\ref{rel1}) and (\ref{rel2}) are related by the following automorphism of the $SL(2,R)$
algebra 
\bea
H \leftrightarrow C \, , \qquad D \to -D \, .
\eea
This is an automorphism of the $SL(2,R)$ algebra which is implemented by a similarity
transformation using the elements 
\bea
g_n=e^{i \pi (n+1/2) H_{\rm osc}}\, ,
\eea
whose action on the generators of $SL(2,R)$ is \footnote{Eq.~(\ref{auto}) is a special case of
\bea
e^{i \theta H_{\rm  osc}} \left( \begin{array}{c} H \\ C \\ D  \end{array}\right) e^{- i \theta H_{\rm osc}}
= \left( \begin{array}{ccc}\cos^2\theta & \sin^2\theta & -\sin\theta \cos\theta  \\
\sin^2\theta & \cos^2\theta & \sin\theta \cos\theta  \\
 \sin 2\theta & -\sin 2\theta & \cos 2\theta
 \end{array}\right)
\left( \begin{array}{c} H \\ C \\ D\end{array}\right)
\eea
The automorphism in Eq.~(\ref{auto}) is obtained for $\sin \theta = \pm 1$.
}
\bea\label{auto}
g_n\left( \begin{array}{c} H \\ C \\ D \end{array}\right)
g_n^{-1}= \left( \begin{array}{c} C \\ H \\ -D \end{array}\right) \, .
\eea
These identities immediately lead to the virial theorems for trapped fermions at the unitary limit
derived in Refs.~\cite{tkt,wc1}. The thermal expectation value of an arbitrary operator,  ${\cal \hat O}$,
is given by
\bea\label{tep}
\langle {\hat O} \rangle =  {\rm Tr}[e^{-\beta (H_{\rm osc}-\mu_+ N_+ -\mu_- N_-)} \hat{O}] \, ,
\eea
where we have included separate chemical potentials, $\mu_+$ and $\mu_-$, for spin up and 
spin down fermions, respectively, so our results can be applied to spin polarized as well as unpolarized gases.
For the expectation value in Eq.~(\ref{tep}),
or the expectation $\hat O$ in an eigenstate of $H_{\rm osc}$,
we have $\langle g_n \hat O g_n^{-1}\rangle = \langle \hat O \rangle$, because 
$[H_{\rm osc},N_\pm]=0$. Therefore, 
\bea\label{mom}
\langle H^n \rangle = \langle  C^n  \rangle \, , \qquad 
\langle   D^{2n+1} \rangle = 0 \, .
\eea
For $n=1$, this implies $\langle H_{\rm osc}\rangle = E_0 = \langle H + C\rangle = 2\langle C\rangle$
which
is the virial theorem first derived in Ref.~\cite{tkt}.
The generalization to arbitrary moments of $C$ in the 
ground state is straightforward:
\bea
\langle \psi_0 | C^n|\psi_0\rangle &=& \langle \psi_0 | C^{n-1}\, C|\psi_0\rangle \nn \\
&=& \langle \psi_0 | C^{n-1}\, \left(\frac{H_{\rm osc}}{2} -\frac{L_+ + L_-}{4}\right) |\psi_0\rangle \nn \\
&=& \frac{E_0}{2}\langle \psi_0 | C^{n-1} |\psi_0\rangle 
+\frac{1}{4}\langle \psi_0 | [L_+,C^{n-1}] |\psi_0\rangle \nn \\
&=&\frac{E_0}{2}\langle \psi_0 | C^{n-1} |\psi_0\rangle 
+\frac{1}{4}\langle \psi_0 | 2(n-1)\,C^{n-1} + [H,C^{n-1}] |\psi_0\rangle \nn \\
&=&\frac{E_0}{2}\langle \psi_0 | C^{n-1} |\psi_0\rangle 
+\frac{1}{4}\langle \psi_0 | 2(n-1)\,C^{n-1} + [H_{\rm osc},C^{n-1}] |\psi_0\rangle \nn \\
&=& \frac{E_0+(n-1)}{2}  \langle \psi_0 | C^{n-1}|\psi_0\rangle \, .
\eea
We have used $L_-|\psi_0\rangle = 0  = \langle \psi_0 | L_+$. 
This simple recursion relation immediately gives all higher moments of the trapping potential
which can be written in closed form as \cite{wc1}
\bea
\langle C^n \rangle = \frac{ \Gamma[E_0+n]}{2^n \Gamma[E_0]} \, .
\eea

This concludes our general discussion of NRCFT.
The main result of this section is the automorphism
of $SL(2,R)$ which relates the two known mappings of the trapped $N$-fermion problem
to problems involving the $N$ fermions in free space. This automorphism provides a 
simple, group theoretical method for deriving virial theorems for both eigenstates and for thermal expectation
values with arbitrary chemical potential for the two spin components.
In the next section of the paper, we will show that the state-operator correspondence 
can be used directly in $d=3$ using the effective field theory of Eq.~(\ref{lag}).  

\section{State-Operator Correspondence in $d=3$}

\subsection{Two fermions}

\begin{figure}[t]
 \begin{center}
 \includegraphics[width=5.0in]{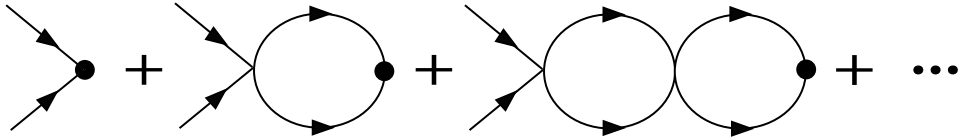}
 \end{center}
 \vskip -0.7cm \caption{Feynman diagrams contributing to the matrix element in Eq.~(\ref{me}).
 The black blob is the operator $\psi \psi$.}
  \label{fig1}
 \end{figure}

The problem of two fermions interacting via short-range interactions in the  presence of an external
harmonic potential is exactly solvable~\cite{Busch:1999}. This solution is reviewed in the Appendix. The
ground state energy of two fermions at the unitary limit in  a harmonic trap is 2, in oscillator units.
In this section, we verify the state-operator correspondence  by evaluating the anomalous dimension of
the composite operator $\psi \psi$ using the NRCFT of Ref.~\cite{Mehen:1999nd}.  We compute the matrix
element
\bea\label{me}
\langle 0| Z_{\psi \psi}(\mu) \psi \psi |\vec{p}, -\vec{p}\,\rangle \, ,
\eea
which is given by the Feynman diagrams in Fig.~\ref{fig1}. The factor $Z_{\psi \psi}(\mu)$ 
is required for composite operator renormalization.
We work in the center of mass frame, where  $E$ is the total
kinetic energy and the momentum of each particle is
 $p =|\vec{p}\,| = \sqrt{ME}$. The diagrams form a geometric series
\bea
\langle 0| Z_{\psi \psi}(\mu) \psi \psi |\vec{p}, -\vec{p}\,\rangle&=&
\frac{Z_{\psi \psi}(\mu) }{1- C_0(\mu) G^0_E(\vec{0},\vec{0}) }\, ,
\eea
where
\bea\label{loop}
G^0_E(\vec{0},\vec{0}) &=& \left(\frac{\mu}{2}\right)^{3-d}\int \frac{d^dl}{(2\pi)^d}
\frac{1}{E - \vec{l}^2/M} \nn \\
&=& -\frac{M}{4\pi}\left(\mu-\sqrt{-p^2+i \epsilon}\right) \, .
\eea
The first line of Eq.~(\ref{loop}) is obtained after evaluating by contour integration the energy integral
in the one-loop bubble graph that is pictured in Fig.~\ref{fig1}. Note that the one-loop graph in the NRCFT
is related to Green's function for the free two-body Hamiltonian:
\bea\label{gff}
G^0_E(\vec{x},\vec{y}) &=&\langle \vec{x}|\frac{1}{E-H_0}|\vec{y}\rangle \nn \\
&=&\left( \frac{\mu}{2}\right)^{3-d}\int \frac{d^dl}{(2\pi)^d}\frac{e^{i \vec{l}\cdot(\vec{x}-\vec{y})}}{E - \vec{l}^2/M}
\, ,
\eea
where $d$ is the number of spatial dimensions and the factor $(\mu/2)^{3-d}$ is 
inserted to give the correct dimensions. We use DR and the PDS scheme~\cite{Kaplan:1998tg} to evaluate
the integral. The integral is linearly dependent on $\mu$ in the PDS scheme, 
reflecting the linear divergence, but $\mu$ independent in the MS scheme.
Keeping the linear $\mu$  dependence is critical for finding the 
correct anomalous dimension for the composite operator $\psi \psi$. 

The result for the matrix element is then 
\bea\label{me2}
\langle 0| Z_{\psi \psi}(\mu) \psi \psi |\vec{p}, -\vec{p}\,\rangle &=&
\frac{Z_{\psi \psi}(\mu)}{1+\frac{M}{4\pi}C_0(\mu)) (\mu + i p) }\nn \\
&=& \frac{4\pi}{M} \left(\frac{Z_{\psi \psi}(\mu)}{C_0(\mu)}\right) \frac{1}{1/a + i p} \, ,
\eea
where we have used Eq.~(\ref{cc}).
The matrix element is $\mu$ independent if $Z_{\psi \psi}(\mu) \propto C_0(\mu)$
and we can fix the constant of proportionality by demanding 
$\langle 0| Z_{\psi \psi}(\mu) \psi \psi |\vec{p}, -\vec{p}\,\rangle = 1$ for $p^2=0$.\footnote{The same
normalization condition is obtained if one requires that the sum of all Feynman diagrams 
yields the same result when evaluated  
in the minimal subtraction (MS) scheme, in which case all loop graphs are finite and $C_0 = 4\pi a/M$.}
Then
\bea
Z_{\psi \psi}(\mu) &=& \frac{M}{4\pi} \frac{C_0(\mu)}{a} \nn\\
&=& \frac{1}{1-\mu\, a}\, .
\eea
The anomalous dimension of the the operator $\psi \psi$ is then given 
by 
\bea
\gamma_{\psi \psi} &=& \mu\frac{d}{d\mu} \, {\rm ln}\, Z_{\psi \psi}(\mu)  \nn \\
&=& \frac{\mu \, a}{1-\mu \, a}\, .
\eea
The effective field theory is a NRCFT when we take the limit $a \to \pm \infty$, and
then $\gamma_{\psi \psi} =-1$. The scaling 
dimension of $\psi \psi$ is the naive dimension, $2 \Delta_\psi =3$, plus the anomalous 
dimension, $\gamma_{\psi \psi}=-1$ so $\Delta_{\psi \psi}=2\Delta_\psi +\gamma_{\psi \psi}=2$, in agreement with the 
state-operator correspondence. Note that one must take the logarithmic derivative with 
respect to $\mu$ at finite $a$, then take the limit $a \to \pm \infty$. If the limit
is taken prior to computing the derivative, then $Z_{\psi \psi}(\mu) =0$. However, this 
is an artifact of the boundary condition that $\langle 0| Z_{\psi \psi}(\mu) \psi \psi |\vec{p},
-\vec{p}\,\rangle = 1$, which is no longer possible when $a =\pm \infty$. If we start with  
Eq.~(\ref{me2}), take the limit $a\to \pm \infty$ we should demand that residue of the 
$1/p$ pole be a $\mu$-independent constant and we again obtain $\gamma_{\psi \psi}=-1$.

Another way of obtaining the scaling dimension of a primary operator, ${\cal O}$, in NRCFT
is to consider the two-point function:
\bea\label{greensf}
G_{\cal O}(\vec{x},t) &=& \langle 0 | {\cal O}(\vec{x},t)\, {\cal O}^\dagger(\vec{0},0)|0\rangle \nn \\
&\propto& \Theta(t) \,t^{-\Delta_{\cal O}}\exp\left( - i N_{\cal O} \frac{\vec{x}^{\,2}}{t}\right) \, ,
\eea
where the second line of Eq.~(\ref{greensf}) follows from scale and Galilean invariance~\cite{Nishida:2007pj}.
Note we assume $N_{\cal O} > 0$. Fourier transforming this Green's function yields
\bea\label{momspace}
\tilde G_{\cal O}(\vec{p},E) 
&=& \int d^d x \, dt \, e^{-i E t + i \vec{p}\cdot\vec{x}} \, G_{\cal O}(\vec{x},t) \nn \\
&\propto& \frac{1}{(E- \frac{p^2}{2N_{\cal O}})^{d/2-\Delta_{\cal O}+1}}
\eea
This formulae can be used once additive renormalizations are carried out.
For example, in the noninteracting theory, for $d=3$,
\bea
\tilde G_{\cal \psi \psi}(\vec{p},E) = -i \frac{M}{4\pi}\left( \mu
-\sqrt{-M E + \frac{p^2}{4} +i\epsilon} \right) 
\eea
After removing the $\mu$ dependence using an additive renormalization, 
we can compare with Eq.~(\ref{momspace}) to obtain $\Delta_{\psi \psi} =3$,
which is the correct answer for a free theory. For the interacting theory,
\bea
\tilde G_{\cal O}(\vec{p},E) &=& \frac{M}{4\pi}\frac{i}{a^2}
\left( \frac{1}{\mu -1/a} + \frac{1}{-\sqrt{-M E +p^2/4 -i \epsilon} +1/a} \right) \, .
\eea
In deriving this result we have included the factor $Z_{\psi \psi}(\mu)$ computed earlier.
The first term can be removed by additive renormalization or else we can remove the
cutoff dependence by taking $\mu \to \infty $. Then, the second term 
then yields $\Delta_{\psi \psi} =2$ when comparing with Eq.~(\ref{momspace})
in the limit $a\to \pm \infty$. 

An alternative formulation of the NRCFT employs a composite field, which in the 
context of nuclear physics is called the dibaryon formalism~\cite{Bedaque:1997qi}. 
This formalism is used most often in three-body calculations.
In the dibaryon formalism one introduces a composite field, $\phi$  
that has the same quantum numbers as $\psi \psi$ and removes the four-fermion interaction 
using a Hubbard-Stratonovich transformation~\cite{Bedaque:1997qi}. 
In this formalism, Eq.~(\ref{momspace}) can be directly compared
to the dibaryon propagator (see, e.g., Eq.~(3) of Ref.~\cite{Bedaque:1997qi}),
in the limit $a\to \pm\infty$, $r_0 \to 0$ and again one finds $\Delta_\phi=\Delta_{\psi \psi} = 2$

It is also interesting to see how the scaling behavior of the two-body wavefunction
dictates the anomalous dimension of the corresponding two-body operator in the field theory.
This sheds further light on the relationship between the results of Ref.~\cite{wc1} and Ref.~\cite{Nishida:2007pj}.
The unrenormalized sum of all graphs in Fig.~\ref{fig1} can be expressed
quantum mechanically as
\bea\label{lswf}
\langle 0 |\psi \psi |\vec{p},-\vec{p} \,\rangle  = \int \frac{d^dq}{(2\pi)^d} 
 \, \langle \vec{q} \,| 1+ \frac{1}{E-H_0} \, T \,|\vec{p}\,\rangle \, ,
\eea
where $T$ is the transition operator that is a solution to the Lippmann-Schwinger equation, 
\bea
T = V + V\frac{1}{E-H_0} \, T \, ,
\eea
and matrix elements of $V$ are $\langle \vec{q}\,|V|\vec{p}\,\rangle = C_0$. Here, we have let the number 
of spatial dimensions, $d$, be arbitrary.
It well known from nonrelativistic quantum mechanics that the exact solution to the scattering wave equation with 
incoming particles with momentum $\vec{p}$ is 
\bea
\chi_{\vec p}(\vec{x}) = \langle \vec{x} \,|1+ \frac{1}{E-H_0} \, T \,|\vec{p}\,\rangle 
\eea
so we can write
\bea
\langle 0 |\psi \psi |\vec{p},-\vec{p} \,\rangle  =  \chi_{\vec p}(0)\, ,
\eea
so the matrix element can be interpreted as two-body wavefunction at the origin.
However, this is divergent for the interacting theory. For two fermions at the unitary limit the two-body wavefunction,
$\chi_{\vec p}(\vec{x})$, is proportional to $r^{2-d}$ for small $r$. Inserting a complete set of state into 
Eq.~(\ref{lswf}) and regulating the expression with a 
hard cutoff in momentum space, we obtain 
\bea
\langle 0 |\psi \psi |\vec{p},-\vec{p} \,\rangle =\chi_{\vec p}(0) = \int^\Lambda \frac{d^dq}{(2\pi)^d} 
\, \int d^d x \, e^{-i \vec{q}\cdot \vec{x}} \,\chi_{\vec p}(\vec{x}) \, .
\eea
This integral is of course divergent. The degree of divergence is determined by the $\vec{x}\to 0$ behavior of 
$\chi_{\vec p}(\vec{x})$ which is independent of $\vec{p}$. It is easily seen that the integral diverges like 
\bea
\int^\Lambda \frac{d^dq}{(2\pi)^d} \int d^d x \, e^{-i \vec{q}\cdot \vec{x}} \frac{1}{r^{d-2}} \sim \Lambda^{d-2} \, .
\eea
If renormalize the matrix element in Eq.~(\ref{lswf}) with a multiplicative factor of 
$Z_{\psi \psi}(\Lambda)$, we must have $Z_{\psi \psi}(\Lambda) \propto \Lambda^{2-d}$ to get a finite answer for the matrix element. 
This leads to $\gamma_{\psi \psi} = 2-d$ which gives for the scaling dimension for $\Delta_{\psi \psi} =
d+ \gamma_{\psi \psi} = 2$, which is the correct answer for arbitrary $d$.

\subsection{Three particles}

The three-body problem in the presence of an external harmonic potential with infinite two-body scattering 
length was solved in Ref.~\cite{wc3}. For any interaction to take place two of the three particles 
must be in an $S$-wave.  Ref.~\cite{wc3} solved the three-body  problem for arbitrary $l$,
where $l$ is the total angular momentum of the three-body system, using
the method of pseudopotentials.  Since the interaction is modeled as zero-range,
the three particles are free except when the coordinates of two of the particles
coincide. The wavefunction is then a solution to the free Schr\"odinger equation
subject to the boundary condition (for arbitrary $a$)
\bea
\label{ppbcs}
\underset{r_{ij}\to 0}{\rm lim}
 \psi(\vec{r}_1,\vec{r}_2,\vec{r}_3) \propto \frac{1}{r_{ij}} -\frac{1}{a} +O(r_{ij}) \qquad  (s_i = -s_j )
\eea
where $r_{ij}\equiv |\vec{r}_i-\vec{r}_j|$, $s_i$ and $s_j$ are the spin quantum numbers of particle $i$ and $j$, 
respectively, and the limit $r_{ij}\to 0$ is taken holding the coordinate of the third 
particle fixed. (For $s_i=s_j$ the wavefunction must vanish as $r_{ij}\to 0$.)
In this paper, we will only consider the case of $l=0$. It would be interesting to extend
the analysis to arbitrary $l$ but that is beyond the scope of this paper.

First we  briefly review the solution obtained in  Ref.~\cite{wc3}. Suppose we choose the spin states so that 
$s_1=s_3=-s_2$. The three-body wavefunction is parameterized as
\bea
\psi(\vec{r}_1,\vec{r}_2,\vec{r}_3) = (1-P_{13})\,\psi_{\rm cm}(\vec{R}_{\rm cm}) \,\frac{1}{r\,\rho} \,\chi(r,\rho) \, ,
\eea
where $R_{\rm cm}$ is the center-of-mass coordinate, $r=|\vec{r}_1-\vec{r}_2|$,  
$\rho = |2\,\vec{r}_3-\vec{r}_1-\vec{r}_2|/\sqrt{3}$,
and the operator $P_{13}$ interchanges $\vec{r}_1$ and $\vec{r}_3$. The wavefunction
for the center-of-mass coordinate, $\psi_{\rm cm}(\vec{R}_{\rm cm})$, is a solution of the simple 
harmonic oscillator Hamiltonian, so $E_{\rm cm} = \omega (2 n + l + 3/2)$.
The function $\chi(r,\rho)$ obeys
the following differential equation
\bea\label{chidiffeq}
\left(\frac{\partial^2}{\partial r^2} +\frac{\partial^2}{\partial \rho^2} 
- \frac{ M^2 \omega^2}{4}(r^2 +\rho^2) + M (E-E_{\rm cm}) \right) \chi(r,\rho) =0 \,  .
\eea
Imposing the boundary conditions in Eq.~(\ref{ppbcs}), and demanding the wavefunction
be finite as $\rho \to 0$, one finds that 
\bea\label{eve}
\frac{\partial}{\partial r} \chi(0,\rho) +\frac{1}{a}\chi(0,\rho) - \frac{4}{\sqrt{3} \rho} \chi\left(\frac{\sqrt{3}}{2}\rho,
\frac{1}{2}\rho\right) =0 \, , \qquad
\chi(r,0) = 0  \, .
\eea
For $a =\pm \infty$, it is possible to solve the boundary condition with a factorized solution,
$\chi(r,\rho) = F_n(R)\phi_n(\alpha)$, where $2 \, R^2=r^2 +\rho^2$, and $\alpha =\arctan(r/\rho)$.
The function $\phi_n(\alpha)$ is determined by
\bea\label{phibcs}
-\frac{\partial^2}{\partial \alpha^2}\phi_n(\alpha) &=& s_{0,n}^2 \, \phi_n(\alpha)\, , \nn \\
 \phi_n(\pi/2) &=&0 \, ,\nn \\
\phi_n^\prime(0)&=&\frac{4}{\sqrt{3}}\phi_n(\pi/3) \, .
\eea
while $F(R)$ obeys the differential equation
\bea\label{Fde}
\left(\frac{\partial^2}{\partial R^2} + \frac{1}{R}\frac{\partial}{\partial R}  
-\frac{s_{0,n}^2}{R^2} - M^2 \omega^2 R^2 +  2 M (E-E_{\rm cm})\right) F(R) = 0
\eea
In addition we must have $\phi_n(0)\neq 0$, so that the residue of the $1/r_{12}$ pole in 
Eq.~(\ref{ppbcs}) is not equal to zero. The first two lines of Eq.~(\ref{phibcs})
are solved by 
\bea\label{phin}
\phi_n(\alpha) \propto \sin \left[\left(\alpha-\frac{\pi}{2}\right) s_{0,n}\right].
\eea
while the third line of Eq.~(\ref{phibcs}) leads to the transcendental equation
for $s_{0,n}$
\bea\label{traneq}
s_{0,n} \cos{\left(\frac{\pi s_{0,n}}{2}\right)}+\frac{4}{\sqrt{3}}
\sin{\left(\frac{\pi s_{0,n}}{6}\right)}
=0\,.
\eea
Note the solutions come in pairs, $s_{0,n} = \pm |s_{0,n}|$.
$s_{0,n}=\pm 2$ is a solution to  Eq.~(\ref{traneq}), however, inspection of Eq.~(\ref{phin})
shows that for $s_{0,n} =\pm 2$ , $\phi_n(0)=0$, which will not satisfy the boundary condition
in Eq.~(\ref{ppbcs}). There are no other integer solutions to Eq.~(\ref{traneq}), and all remaining 
solutions to Eq.~(\ref{traneq}) give nontrivial solutions to the three body-problem. 
Numerical values of the five smallest values of $|s_{0,n}| $ are $2.16622, 5.12735,7.11448, 8.83225, 11.06273$.
The numbers $s_{0,n}$ determine the energy eigenvalues via Eq.~(\ref{Fde}). The solutions
of Eq.~(\ref{Fde}) are~\cite{wc3}
\bea\label{Fsln}
F_n(R) \propto R^{s_{0,n}} \, e^{-R^2 \,M \omega/2} \,L_q^{(s_{0,n})}(R^2\, M\,\omega) \, ,
\eea
where $L_q^{(s_{0,n})}$ is a generalized Laguerre polynomial, and the energy eigenvalue is $E=E_{\rm cm} +\omega (s_{0,n} +1 +2 q)$. 
Note that for the wavefunction to be square integrable, we must have $s_{0,n}$ positive in Eq.~(\ref{Fsln}).
The dependence on the quantum number $q$ shows that for each $s_{0,n}$ there is an infinite  tower of evenly
spaced states whose energies are separated by $2\, \omega$, as expected 
from the  $SL(2,R)$ algebra.

At this point we would like to demonstrate the correspondence between trapped eigenstates and zero-energy, 
scale-invariant eigenfunctions of the free Hamiltonian. To find these states,
we can choose the same variables, $R_{\rm cm}, R$, and $\alpha$, which were used to solve the 
trapped three-body problem. The function $\psi_{\rm  cm}(\vec{R}_{\rm cm})$ is now a solution
to the free particle Schr\"odinger equation, $\psi_{\rm  cm}(\vec{R}_{\rm cm}) \propto
e^{- i \vec{P}_{\rm cm} \cdot \vec{R}_{\rm cm}}$, and we should take 
$\vec{P}_{\rm cm}= 0$ to obtain a zero-energy state. The eigenvalue equations for $\phi_n(\alpha)$
are still Eq.~(\ref{eve}), and $F_n(R)$ obeys Eq.~(\ref{Fde}) with $\omega =E=E_{\rm cm}=0$.
Thus, the solution for the zero-energy scale-invariant wavefunction has $F_n(R)\propto R^{s_{0,n}}$
and the zero-energy, scale-invariant solution to the three-body equation is
\bea
\psi_\nu(\vec{r}_1,\vec{r}_2,\vec{r}_3) &\propto& (1-P_{13}) \, \frac{1}{r \,\rho} R^{s_{0,n}} \phi_n(\alpha) \nn \\
&=& (1-P_{13}) \,  R^{s_{0,n}-2} \frac{\phi_n(\alpha)}{\sin(2\alpha)} \, .
\eea
Clearly, the scaling exponent for this state is $\nu=s_{0,n} -2$, so Eq.~(\ref{enev}) tells
us that the energy of the ground state of the infinite tower of states  
is $E=\omega(5/2 +s_{0,n})$, in agreement with the result obtained by 
direct solution of the three-body equations.

Now we would like to see how the  effective field theory reproduces these results. We will
show that the effective field theory allows one to derive a bound state equation which 
exhibits scaling solutions whose scaling exponents yield energy eigenvalues via
the correspondence of Ref.~\cite{wc1}. Then we study how the state-operator correspondence 
can be used directly in three dimensions by calculating the anomalous dimension of an operator
in the NRCFT  and verifying that it reproduces the  known result for the lowest energy state
of three trapped particles in an $S$-wave.

In applications of effective field theory to three-body problems it has been found useful to employ the 
dibaryon formalism discussed earlier~\cite{Bedaque:1998kg}. In the present context the composite field should be called a 
difermion, which we will denote $\phi$, which has the same quantum numbers as $\psi \psi$.  A 
Hubbard-Stratonovich transformation is used to trade the contact interaction 
in Eq.~(\ref{lag}) for $\phi^\dagger \psi \psi$ and $\psi^\dagger \psi^\dagger \phi$ couplings.
Loops of $\psi \psi$ contributing to the $\phi$ self-energies are summed to all orders to obtain
the $\phi$ propagator. We refer readers to Ref.~\cite{Bedaque:1998kg} for details on this procedure.

\begin{figure}[t]
 \begin{center}
 \includegraphics[width=6.0in]{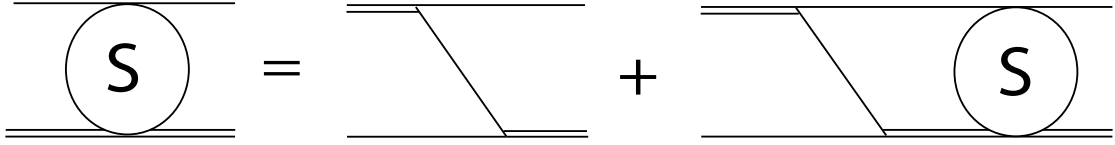}
 \end{center}
 \vskip -0.7cm \caption{Integral equation for $\phi \psi$ scattering.}
  \label{tbint}
 \end{figure}

The scattering of $\phi$ and $\psi$ proceeds via an infinite number of ladder-like 
diagrams. These can be resummed using a one-dimensional integral equation, which is pictured
in Fig.~\ref{tbint}. Double lines are $\phi$ propagators and single lines are $\psi$ propagators.
Evaluating the diagrams in Fig.~\ref{tbint} and projecting onto the $S$-wave
yields the half-off-shell integral equation 
\bea\label{Sinteq}
 S(p,k) &=& -\frac{M}{2 p k} \log\left(\frac{p^2 +p k +k^2 -M E}{p^2 -p k +k^2 -ME}\right) \\
&-& \frac{1}{\pi}\int_0^\infty dq \frac{1}{pq}
\log\left(\frac{p^2 +p q +q^2 -M E}{p^2 -p q +q^2 -ME}\right)
\frac{q^2}{-1/a + \sqrt{3q^2/4 -M E}} \, S(q,k) \, .  \nn
\eea
Here the momentum $k$ is on-shell, $3 k^2/4= M E + 1/a^2$, and the momenta $p$ and $q$
are off-shell. Note that $S(p,k)$ corresponds to the sum of ladder diagrams and does not
include the LSZ factors required to obtain the on-shell amplitude when $p=k$, nor
is it normalized to give the three-body scattering length for $p=k=0$. 
Apart from these factors, 
the equation obtained here for $\phi\, \psi$ scattering is identical to that obtained 
in Ref.~\cite{Bedaque:1998kg} for three nucleons in the $J=3/2$ state of nucleons. 
Ref.~\cite{Bedaque:1998kg} defines a half-off-shell amplitude, 
$a(p)$, which is normalized to the three-body scattering length, $a_3=a(p=k)$.
The function $S(p,k)$ is related to the function $a(p)$ of Ref.~\cite{Bedaque:1998kg}
by
\bea
S(p,k) = -\frac{3 M}{8}\frac{a(p)}{1/a+\sqrt{3 p^2/4 - M E}} \, .
\eea 

It is straightforward to reproduce the results of Ref.~\cite{wc3} using the NRCFT
and the mapping of Ref.~\cite{wc1}. To find a zero-energy, scale-invariant eigenstate
of the free space problem we can consider the equation for three-body bound states
in the limit $a\to \pm \infty$ and $E=0$. The bound state equation is obtained
from Eq.~(\ref{Sinteq}) by dropping the inhomogeneous term in the integral equation.
In the limit $E\to 0$ and $a\to \pm \infty$, the bound state equation becomes
\bea\label{bse}
S(p,0) = -\frac{2}{\pi \sqrt{3}}\int_0^\infty \frac{dq}{p}
\log\left(\frac{p^2 +p q +q^2}{p^2 -p q +q^2}\right) S(q,0)\, .
\eea
Then one looks for solutions of the form $S(p,0)\propto p^{-s_{0,n}-1}$, which is
possible if $s_{0,n}$ satisfies Eq.~(\ref{traneq})~\cite{Bedaque:1998kg}.
The integral equation for $S(p,k)$ is finite and does not require renormalization.
In order for the second diagram on the right hand side of the integral equation 
in Fig.~\ref{tbint} to converge for large $q$, $S(q,0)$ must vanish as $q \to \infty$,
which then leads to $S(p,0) \propto p^{-|s_{0,n}|-1}$. To see how the $p \to \infty$
behavior of $S(p,0)$ is related to the scaling behavior of the many-body 
wavefunction, we recall the three-body position space wavefunctions can be obtained from $S(p,k)$ using
the following transform~\cite{Bedaque:1998qu}
\bea\label{transform}
\chi(r,\rho) = \int^\infty_0 dp \,S(p,k) \frac{p \sin\left(\sqrt{\frac{3}{4}}p\, \rho\right)}{-1/a+ \sqrt{\frac{3}{4}p^2-M E}}
e^{-r \sqrt{\frac{3}{4}p^2-M E}} \, .
\eea
It is straightforward to show that the function $\chi(r,\rho)$ obeys Eqs.~(\ref{chidiffeq},\ref{eve}).
Eq.~(\ref{chidiffeq}) and the second boundary condition in Eq.~(\ref{eve}) follow directly from the 
definition in Eq.~(\ref{transform}), while the first boundary condition in Eq.~(\ref{eve})  
can be obtained using the integral equation for $S(p,k)$. Taking the limit $E = 0$, $a =\pm \infty$,
and inserting the the asymptotic solution for $S(p,0)$, we find
\bea
\chi(r,\rho) &\propto & \int_0^\infty dp \,p^{-s_{0,n}-1}\, \sin\left( \sqrt{\frac{3}{2}} \, p \, R \, \cos\alpha\right) 
e^{-\sqrt{3/2} \, p \, R \, \sin\alpha} \nn \\
&\propto& R^{s_{0,n}} \,\sin\left[ s_{0,n}\left(\alpha-\frac{\pi}{2}\right)\right]\, .
\eea
which is the correct form of the zero-energy, scale-invariant solution.

\begin{figure}[t]
 \begin{center}
 \includegraphics[width=3.0in]{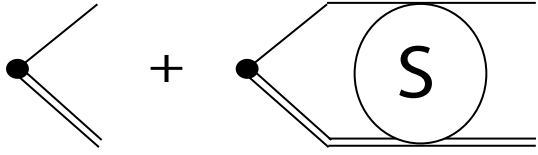}
 \end{center}
 \vskip -0.7cm \caption{Diagrams contributing to the renormalization of the operators ${\cal O}_1$.}
  \label{ppren}
 \end{figure}

Finally we wish to understand the state-operator correspondence for the case of the 
three trapped fermions in an $S$-wave. 
As an example, we show how  the state-operator correspondence can be used 
to calculate the lowest energy state of three harmonically trapped fermions in an $S$-wave.
We compute the scaling dimension of the operator
\bea
{\cal O}_1 = \phi \, i  \frac{\overleftrightarrow{\partial}}{\partial t} \,  \psi \, .
\eea 
Operators that are a total time or space derivative are not primary, so ${\cal O}_1$
is the unique primary operator with one time derivative. 
$S$-wave operators with two space-derivatives can be put in the form $\phi \nabla^2\psi$ after
integration by parts, and are therefore equivalent by the equations of motion for $\psi$. 
Therefore, in the noninteracting theory, where $\phi$ has dimension 3,  ${\cal O}_1$ is the unique 
operator with naive dimension $13/2$.  Note that the lowest energy state of three noninteracting
fermions in an $S$-wave has energy $13/2 \, \omega$, so the naive scaling dimension is consistent 
with the state-operator correspondence for the free theory.
We compute the matrix element $\langle 0| Z_1(\Lambda) \,{\cal O}_1 |\vec{p}, -\vec{p}\,\rangle$. 
The diagrams that contribute to this matrix element are pictured in Fig.~\ref{ppren},
which shows a tree-level graph and another graph which includes the half-off-shell 
 amplitude $S(p,k)$. The off-shell legs of $S(p,k)$ are contracted with the operator ${\cal O}_1$
 to form a loop. This graph sums all loop corrections to the matrix element
 $\langle 0| Z_1(\Lambda) \,{\cal O}_1|\vec{p}, -\vec{p}\,\rangle$.
The sum of all diagrams contributing to the renormalization of  ${\cal O}_1$ 
in the limit $E, 1/a = 0$  is given by
\bea\label{tbme}
Z_1(\Lambda) \int \frac{d^4p}{(2\pi)^4} \,
\frac{i}{p_0 -\frac{p^2}{2 M} + i \epsilon}\,
\frac{4\pi}{M}
\frac{-i}{\sqrt{M p_0  + p^2/4- i \epsilon}} \, 2 p_0 \, i S(p,0)  \, \nn \\
= Z_1(\Lambda) \, \frac{4}{\sqrt{3}\,\pi M^2}\int_0^\Lambda \, dp \, p^3\, S(p,0) \,.
\eea
We have included the factor $Z_1(\Lambda)$ for composite operator renormalization.
A cutoff on the virtual loop momentum is used to regulate the loop integral.
The asymptotic form of $S(p,0)$ is determined by Eq.~(\ref{bse}), so in general
we have
\bea\label{msum}
S(p,0) = \sum_{m} c_m p^{-|s_{0,m}|-1} \, 
\eea
where the coefficients in the expansion, $c_m$, must be determined numerically from
the solution to the full integral equation. Obviously the large $p$ behavior 
is dominated by the smallest values of $m$ in Eq.~(\ref{msum}). For the operator ${\cal O}_1$,
the only divergent contribution comes from the term $m=1$, $|s_{0,1}|=2.16622$. 
All other terms in Eq.~(\ref{msum}) give a UV finite contribution to Eq.~(\ref{tbme}).
The integral on the right hand side of Eq.~(\ref{tbme}) diverges as $\Lambda^{3-|s_{0,1}|}$, so 
\bea
\gamma_1 = \Lambda \frac{d}{d\Lambda} Z_1(\Lambda)
=|s_{0,1}|-3 \, .
\eea
The scaling 
dimensions of $\phi$ and $\psi$ are $\Delta_\phi=2$ and $\Delta_\psi =3/2$, as discussed in the 
previous section, and the time derivatives add $2$ to the naive dimension
of ${\cal O}_1$. 
Adding the anomalous dimension, $\gamma_{\phi \psi}$, we find the scaling 
dimension, $\Delta_1 = 5/2 + |s_{0,1}|$, which, via the state-operator correspondence, is also in agreement with the result
for the lowest energy state of three harmonically trapped fermions in an $S$-wave~\cite{wc3}.

We should not consider the operator $\phi \psi$. 
The analog of this operator in the formulation of the theory without a difermion field 
would be $(\psi \psi) \psi = \epsilon^{\alpha \beta} \psi_\alpha \psi_\beta  \psi$ which
is not allowed because of Fermi statistics. A local operator which creates (or annihilates)
three fermions at a point must have derivatives acting on at least one of the fermion fields. Therefore
the  operator $\phi \psi$  which seems allowed if $\phi$ is treated as a boson, must be excluded
from consideration when classifying local operators in the NRCFT. This can also be seen from
the state-operator correspondence for the noninteracting theory. The naive scaling dimension
of $\phi \psi$ is $9/2$ in this case, but there is no state of three trapped fermions
with energy $9/2 \, \omega$. So clearly one obtains a contradiction with the state-operator 
correspondence if $\phi \psi $ is allowed.

Note that the true ground state of three fermions at the unitary limit has $l=1$.  It would be
interesting to derive the transcendental equations analogous to Eq.~(\ref{traneq}) for $l \neq 0$ from
the integral equations for\ scattering in higher partial waves derived in Ref.~\cite{Gabbiani:1999yv}.
These should give the numbers $s_{l,n}$ that determine the energy eigenvalues of three
fermions  in higher partial waves~\cite{wc3}. Another problem is to determine operators that correspond
to states with energy eigenvalues $E = 5/2 + |s_{0,n}|$, $n \geq 2$. These come from operators with two
or more time derivatives or four or more spatial derivatives, or mixed time and space derivatives. In
the equations analogous to Eq.~(\ref{tbme}), these operators will lead to more factors of $p_0$ or
$p^2$ which will make the integral more divergent.  This leads to more terms in the sum in
Eq.~(\ref{msum}) contributing to the anomalous dimension. It should be possible to find a basis of
operators  in which  the anomalous dimensions are given by the $s_{0,n}$ that are solutions to
Eq.~(\ref{traneq}).

\section{Conclusions}

We have studied the problem of  fermions with infinite two-body scattering length confined in harmonic
traps. The $N$-body problem can be  mapped to problems involving $N$ fermions in the absence of an
external potential. One approach is to map solutions of the trapped problem to zero-energy,
scale-invariant  solutions to the Schr\"odinger equation in free space~\cite{wc3}. Another approach is
to relate the energy levels of $N$-fermion states to the scaling dimensions of  primary operators in an
NRCFT~\cite{Nishida:2007pj}. In this paper, we have shown that these two mappings are related by an
automorphism of the $SL(2,R)$ conformal algebra of the NRCFT. This automorphism  interchanges the
internal Hamiltonian of the NRCFT with the harmonic trapping potential. This provides a simple, group theoretical way of
deriving virial theorems for trapped Fermi gases at the  unitary limit. The virial theorems apply for
energy eigenstates as well as thermal ensembles and  hold for both spin polarized and unpolarized gases.

One goal of this paper was to apply the state-operator correspondence~\cite{Nishida:2007pj} directly in the three spatial
dimensions ($d=3$), which  is clearly the most important case. In Ref.~\cite{Nishida:2007pj}, the
state-operator correspondence was combined with $\epsilon$ expansions about $d=2$  and $d=4$ to do
perturbative calculations of the energy levels. We sought to apply the  state-operator correspondence
directly in three dimensions using the NRCFT of Eq.~(\ref{lag}).  This is clearly more difficult because
analytic results are only available for two fermions. For two fermions we showed how to use the
state-operator correspondence to calculate the energy levels of two trapped fermions at the unitary
limit. For three fermions, the NRCFT gives an integral  equation for $\phi \psi$ scattering which can
used to find the zero-energy, scale-invariant eigenfunctions which can be used to find the
eigenfunctions of the three trapped fermions via the correspondence of Ref.~\cite{wc3}. We showed
how to use the state-operator correspondence to  calculate the energy  of the lowest 
energy $S$-wave three-fermion trapped state. It would be interesting to extend application
of the state-operator correspondence to all eigenstates of the trapped three fermion problem.

Since the problems of two and three trapped fermions in the unitarity limit can be solved using quantum
mechanics  and the pseudopotential boundary conditions of Eq.~(\ref{ppbcs}), an important question is
whether the mappings of the trapped fermion problems to free space problems will be useful for
obtaining new results. The virial theorems~\cite{wc1} are an example of results that the conformal
symmetry  of the NRCFT can provide in the absence of an exact solution of the quantum mechanics
problem. It would be interesting if the integral equations of the effective field theory  could be used
to calculate corrections to energy levels from a finite scattering length,  or obtain new results for
problems with four or more fermions at the unitary limit. It would also be interesting if $SL(2,R)$
invariance can be used to obtain  information about correlation functions of  two-point functions of
primary operators in the eigenstates of harmonically trapped fermions. For example, if $SL(2,R)$ 
invariance provides interesting constraints on correlation
functions like Eqs.~(\ref{greensf},\ref{momspace}) with the vacuum replaced by the ground state of $N$
trapped fermions,  one could perhaps learn something about the low lying excitations of the ground state 
of a trapped gas of cold atoms at the unitary limit.

\acknowledgments 

We thank Iain Stewart and Paolo Bedaque for discussions on topics related to this paper, We thank Sean Fleming, Bira van
Kolck, Berndt Mueller and Dam Son for comments on an earlier draft of this manuscript. We also thank Yusuke Nishida for
sharing unpublished notes on calculations relevant to Section III of this paper. This work was supported in part by the
Department of Energy under grant numbers DE-FG02-05ER41368, DE-FG02-05ER41376, and DE-AC05-84ER40150.

\appendix
\section{Two Fermion is a Harmonic Trap in EFT}

In this appendix we solve the problem of two-particles interacting via short 
range forces in the presence of a harmonic potential. This problem was first 
solved in Ref.~\cite{Busch:1999} and is typically analyzed using the method 
of pseudopotentials, see e.g. Refs.~\cite{Tiesinga:2000, Blume:2002, Block:2002,
Bolda:2002, Stock:2005, Idziaszek:2006a, Idziaszek:2006b}. Here we solve it by 
evaluating the two-particle Green's function.

Consider the Green's function
\bea
G^{(0)}_{E_{\rm tot}}(\vec{x}_3,\vec{x}_4;\vec{x}_1,\vec{x}_2)
 &=&\langle \vec{x}_3,\vec{x}_4|\frac{1}{E_{\rm tot}-H^{(0)}}|\vec{x}_1,\vec{x}_2\rangle \, ,
\eea
where $H(H^0$) corresponds to the interacting (noninteracting) Hamiltonian.
$G_{E_{\rm tot}}(\vec{x}_3,\vec{x}_4;\vec{x}_1,\vec{x}_2)$ obeys the integral equation
\bea
G_{E_{\rm tot}}(\vec{x}_3,\vec{x}_4;\vec{x}_1,\vec{x}_2) &=&
G^0_{E_{\rm tot}}(\vec{x}_3,\vec{x}_4;\vec{x}_1,\vec{x}_2) \nn\\
&&+\, C_0(\mu) \, \int d^Dy \,  G^0_{E_{\rm tot}}(\vec{x}_3,\vec{x}_4;\vec{y},\vec{y}\,) \,
G_{E_{\rm tot}}(\vec{y},\vec{y};\vec{x}_1,\vec{x}_2) \, .
\eea
This equation can be derived  in quantum mechanics using a delta-function potential with coefficient $C_0(\mu)$
or from the Feynman diagrams of the field theory in Eq.~(\ref{lag}) in position space.
It is helpful to go to center of mass coordinates 
\bea
\vec{x}_{1,2} = \vec{R} \pm \frac{1}{2}\vec{r} \qquad 
\vec{x}_{3,4} = \vec{R}^{\,\prime} \pm \frac{1}{2}\vec{r}^{\,\prime} \, ,
\eea
because the Hamiltonian factorizes in these coordinates.  The noninteracting Green's
function is given by
\bea
G^{0}_{E_{\rm tot}}(\vec{x}_3,\vec{x}_4;\vec{x}_1,\vec{x}_2)
= \sum_{\vec{n},\vec{m}} \frac{
\psi_{\vec{m}}(\vec{R}^{\, \prime}) \, \phi^0_{\vec{n}}(\vec{r}^{\, \prime}) \,\psi_{\vec{m}}(\vec{R}\,)\, \phi^0_{\vec{n}}(\vec{r}\,)
}{E_{\rm tot}- E(\vec{n})-E(\vec{m})} \, .
\eea
Here $\vec{n} = (n_x,n_y,n_z)$ and $\vec{m} = (m_x,m_y,m_z)$. The $\psi_{\vec{m}}(\vec{R})$
are eigenfunctions of the $H_{\rm CM}$ with energy $E(\vec{m})$, and $\phi^0_{\vec{n}}(\vec{R})$
are eigenfunctions of the noninteracting $H_{\rm rel}$   with energy $E(\vec{m})$.
The interacting Green's function $G^0_{E_{\rm tot}}$ has the same form as $G_{E_{\rm tot}}$
with $\phi^0_{\vec{n}}(\vec{R})$ replaced by eigenfunctions of $H_{\rm rel}$, $\phi_{\vec{n}}(\vec{R})$.
Since $H_{\rm cm}$ is the same in either case, the
 $\psi_{\vec{m}}(\vec{R})$ are common to  $G^0_{E_{\rm tot}}(\vec{x}_3,\vec{x}_4;\vec{x}_1,\vec{x}_2)$ and 
$G_{E_{\rm tot}}(\vec{x}_3,\vec{x}_4;\vec{x}_1,\vec{x}_2)$, we can project 
onto an energy eigenstate of $H_{\rm CM}$. Defining
\bea
G^{(0)}_E(\vec{r}^{\, \prime},\vec{r}\,) \equiv \int d^d\vec{R} \, d^d\vec{R}^{\,\prime} \, 
 \psi_{\vec{m}}(\vec{R}\,) \, \psi_{\vec{m}}(\vec{R}^{\, \prime}) \,
G^{(0)}_{E_{\rm tot}}(\vec{x}_3,\vec{x}_4;\vec{x}_1,\vec{x}_2) \, ,
\eea
where $E =E_{\rm tot} - E(\vec{m})$ and $E(\vec{m})$ is an eigenvalue of $H_{\rm cm}$, 
we find that $G_E(\vec{r}^{\,\prime},\vec{r})$  obeys the equation
\bea\label{inteq}
G_E(\vec{r}^{\, \prime},\vec{r}\,) = G^0_E(\vec{r}^{\, \prime},\vec{r}\,) 
+ C_0(\mu) \, G^0_E(\vec{r}^{\, \prime},\vec{0}) \,G_E(\vec{0},\vec{r}\,) \, .
\eea
The notation is similar to that used in Eq.~(\ref{gff}), however here $H_0$ is the simple harmonic oscillator Hamiltonian.
Setting $\vec{r}^{\,\prime}=\vec{0}$ we obtain
\bea
G_E(\vec{0},\vec{r}\,) =\frac{G^0_E(\vec{0},\vec{r})}{1-C_0(\mu) \,G^0_E(\vec{0},\vec{0})} \, .
\eea
The poles of this expression are solutions to 
\bea\label{pole}
\frac{1}{C_0(\mu)} - G^0_E(\vec{0},\vec{0}) = 0 \, ,
\eea
where $G^0_E(\vec{0},\vec{0}\,)$ is the Green's function for the simple harmonic oscillator, and 
is given by 
\bea
G^0_E(\vec{0},\vec{0}\,) &=&-\int_0^\infty dt \, \langle \vec{0} | e^{(E-H_0) t}| \vec{0}\, \rangle \nn \\
&=& \int_0^\infty dt \,e^{E t}\left(\frac{M \omega}{4\pi \sinh(\omega t)}\right)^{d/2} \, .
\eea
A transcendental equation similar to Eq.~(\ref{pole}) but with 
a different regulator was obtained in Ref.~\cite{SBvKV}.
This integral is evaluated for negative $E-E_0$, where $E_0$ is the ground state of the oscillator,
using dimensional regularization:
\bea
\int_0^\infty dt \frac{e^{-a t}}{(\sinh t)^{d/2}} &=& 2^{d/2-1}\int_0^1 du \, u^{a/2+d/4-1} \, (1-u)^{-d/2} \nn \\
&=& 2^{d/2-1}\frac{\Gamma[1-\frac{d}{2}]\Gamma[\frac{a}{2}+\frac{d}{4}]}{\Gamma[1-\frac{d}{4}+\frac{a}{2}]} \, .
\eea
Analytically continuing the integral from negative to positive $E$ we find, for arbitrary $d$,
\bea
G^0_E(\vec{0},\vec{0}\,) &=& -\left(\frac{M}{4\pi}\right)^{d/2}(2\omega)^{d/2-1} 
\frac{\Gamma[1-\frac{d}{2}]\Gamma[-\frac{E}{2\omega}+\frac{d}{4}]}{\Gamma[1-\frac{d}{4}-\frac{E}{2\omega}]} \, . \nn \\
\eea
Just like the $G^0_E(\vec{0},\vec{0}\,)$ in the absence of the oscillator potential, this integral
is linear divergent, but finite if evaluated using dimensional regularization. The integral is defined
exactly as in the free space theory, multiplying the integral by 
$(\mu/2)^{3-d}$ and subtracting the pole at $d=2$. We find
\bea
G^0_E(\vec{0},\vec{0}\,) &=& \frac{M}{4\pi} \left( -\mu 
+ \sqrt{2 M \omega}\, \frac{\Gamma[\frac{3}{4}-\frac{E}{2 \omega}]}{\Gamma[\frac{1}{4}-\frac{E}{2 \omega}]}\right) \, .
\eea
Therefore we find the poles of the Green's function are located at,
\bea\label{traneq2}
0&=&\frac{1}{C_0(\mu)}-G^0_E(\vec{0},\vec{0}) \nn \\
&=& \frac{M}{4\pi}\left( \frac{1}{a} - \sqrt{2 M \omega}\,
\frac{\Gamma[\frac{3}{4}-\frac{E}{2 \omega}]}{\Gamma[\frac{1}{4}-\frac{E}{2 \omega}]}\right) \, ,
\eea
which is the transcendental equation first derived in Ref.~\cite{Busch:1999}. This result is easily
generalized to include effective range corrections. Effective range corrections and higher order terms in
the effective range expansion can be incorporated using higher dimension operators with derivatives. 
We can choose a basis where each higher dimension operator contributes a factor of $C_{2n}(M E)^n$
to the tree level scattering amplitude, see Ref.~\cite{Beane:2000fi} for more details. The scattering amplitude is
in the absence of an external potential is
\bea
{\cal A}&=&\frac{-1}{\left(\sum_n C_{2n}(\mu)(M E)^n\right)^{-1}+\frac{M}{4\pi}(\mu +i p)}\nn \\
&=&\frac{4\pi}{M}\frac{1}{p \cot \delta(E) - ip}
\eea
Including the higher derivative operators in the Eq.~(\ref{inteq}) for the Green's function,
one finds the the formula in  Eq.~(\ref{traneq2}) becomes
\bea\label{effrangecorr}
0&=&\frac{1}{\sum_n C_{2n}(\mu) (M E)^n}-G^0_E(\vec{0},\vec{0}\,) \nn \\
&=& \frac{M}{4\pi}\left(-p\cot{\delta(E}) - \sqrt{2 m \omega} \,
\frac{\Gamma[\frac{3}{4}-\frac{E}{2 \omega}]}{\Gamma[\frac{1}{4}-\frac{E}{2 \omega}]}\right) \, ,
\eea 
In Ref.~\cite{Tiesinga:2000} it was pointed out that Eq.~(\ref{traneq2}) receives
significant corrections when $a \sqrt{2 M\omega} = a/a_{\rm osc} \geq 1$. Later, it was shown~\cite{Blume:2002,
Block:2002,Bolda:2002}
showed that reliable results could be obtained by making the  substitution
\bea
\frac{1}{a} \to -p \cot \delta(E)\, .
\eea
This substitution was called the ``effective-scattering length model", which 
we see here can be derived in a straightforward way using effective field theory.

\end{document}